\documentclass[twocolumn,aps,prb]{revtex4}%
\usepackage{amsmath}
\usepackage{amsfonts}
\usepackage{dsfont}
\usepackage{amssymb}
\usepackage{graphicx}
\usepackage{leftidx}
\usepackage[colorlinks,  linkcolor=blue, anchorcolor=blue,citecolor=blue,
urlcolor=blue ,dvipdfmx]{hyperref}
\usepackage{txfonts}%
\DeclareMathAlphabet{\mathcal}{OMS}{cmsy}{m}{n}

\begin{document}

\title{Coexistence of New Fermions in Topological Semimetal TaS}
\author{Jian-Peng Sun$^{1}$, Dong Zhang$^{1}$\footnote{zhangdong@semi.ac.cn}, Kai Chang$^{1}$\footnote{kchang@semi.ac.cn}}
\affiliation{$^{1}$SKLSM, Institute of Semiconductors, Chinese Academy of Sciences,
P.O. Box 912, 100083, Beijing, China}
\begin{abstract}
We theoretically propose that, the single crystal formed TaS is a new type of topological semimetal, hosting ring-shaped gapless nodal lines and triply degenerate points (TDPs) in the absence of spin-orbit coupling (SOC). In the presence of SOC, the each TDP splits into four TDPs along the high symmetric line in the momentum space, and  one of the nodal ring remains closed due to the protection of the mirror reflection symmetry, while another nodal ring is fully gapped and transforms into six pairs of Weyl points (WPs) carrying opposite chirality. The electronic structures of the projected surfaces are also discussed, the unique Fermi arcs are observed and the chirality remains or vanishes depending on the projection directions. On the (010) projected surface, one may observe a Lifshitz transition. The new type of topological semimetal TaS is stable and experimentally achievable, and the coexistence of topological nodal lines, WPs and TDPs
states in TaS makes it a potential candidate to study the interplay between different types of topological fermions.
\end{abstract}

\pacs{73.21.La, 73.22.Dj, 73.22.Gk, 73.20.At}
\maketitle

\section{INTRODUCTION}
Topological semimetals (TSMs) have stimulated tremendous interests in recent years, not only for they are metallic mimic of the topological insulators \cite{RevModPhys.82.3045,1.3293411,nature08916,RevModPhys.83.1057}, but also for they provide fertile platforms to realize unique particles in high-energy particle physics. Although the searching for novel fermions in condensed matters can stem back to 1937 \cite{PhysRev.52.365}, C. Herring proposed accidental two-fold degeneracy of bands crossing in three-dimensional (3D) lattice and found it robust even without any symmetry protection within Weyl function. And later G. Volovik proposed Weyl fermions in superfluid $^3$He-A \cite{PMC26832}. Till 2011, X. Wan et al. proposed the first specific compounds $\text{Re}_{2}\text{Ir}_{2}\text{O}_{7}$ (Re=rare earth element)
\cite{PhysRevB.83.205101} to be Weyl semimetal. After that, abundant TSMs candidates have been proposed.

Current TSMs can be classified as Weyl semimetals,
Dirac semimetals and node-line semimetals according to the
different band crossing points at the Fermi level and the mechanisms protecting them. To be more specifically,
a Dirac semimetal \cite{PhysRevLett.108.140405,PhysRevLett.115.126803,
PhysRevB.85.195320,PhysRevLett.116.186402} is characterized by
hosting Dirac points of linear crossing of two doubly degenerate bands in momentum
space near the Fermi level. The Dirac points are protected by certain crystalline symmetry, as a result, which makes the quadruple degenerate points located either
at the high-symmetry point or along high-symmetry lines.
The node-line semimetal \cite{PhysRevB.84.235126,PhysRevLett.115.036807,
PhysRevLett.115.036806,PhysRevB.92.081201,ncomms10556,PhysRevB.93.121113,PhysRevB.93.205132,PhysRevB.94.121108,
PhysRevB.93.201114,PhysRevB.94.155121,027102} is another type of TSM where the valence and conduction
bands cross along one-dimensional (1D) lines in momentum space
which results in forming a ring-shaped nodal line.
Different from the Dirac and node-line semimetals, the Weyl semimetal
\cite{PhysRevB.83.205101,PhysRevLett.107.127205,PhysRevB.85.035103,
PhysRevX.5.011029,Xu613,PhysRevB.92.161107,PhysRevLett.117.066402,
PhysRevB.94.161401,nphys3871,PhysRevX.6.031021} is classfied by possessing Weyl points (WPs) that are point-like band crossings of two nondegenerate bands with linear dispersion near the Fermi level. Weyl semimetal does not require any protection from the
crystalline symmetry other than lattice translation, which makes WPs
possibly located at generic $\mathbf{k}$ points. These WPs with opposite chirality represent sources or sinks of Berry curvature and have to appear in pairs. They are associated with the topological Chern number $\mathcal{C}=\pm 1$
and positive (negative) Chern number is equivalent to a source (sink) of Berry curvature.
Both Weyl and Dirac semimetals possess Fermi surfaces consisting
of a few crossing points in the Brillouin zone (BZ). However, the Fermi surface of a node-line semimetal is a
closed ring-shaped nodal line arising from the valence and
conduction bands crossing along the specific crystllagraphic
orientations in the BZ.

Besides the above-mentioned topological semimetals,
another class of topological semimetal materials characterized
by three- or sixfold band crossings have been proposed and
are named as ``new fermion" by Bradlyn $et$ $al$ \cite{Bradlynaaf5037}.
Very recently, the TSMs with triply degenerate points (TDPs) have been predicted theoretically in some materials with hexagonal
lattice structure \cite{PhysRevX.6.031003,PhysRevB.94.165201} and
the CuPt-ordered alloys $\text{InAs}_{0.5}\text{Sb}_{0.5}$ \cite{PhysRevLett.117.076403}.
In the band structures of these materials, both one- and two-dimensional representations are allowed along a certain high-symmetry axis, which makes it possible to generate band crossing between a doubly degenerate band and a nondegenerate band near the Fermi level and form a TDP \cite{PhysRevB.87.045202}. This new type of three-component fermions can be regarded as the ``intermediate state" between the four-component Dirac and the two-component Weyl fermions.

As the new fermions emerge, the interplay among the fermions reveals underlying physics. For example, the Dirac fermions can split into one pair of Weyl fermions with opposite chirality by breaking spatial inversion or time-reversal symmetry \cite{PhysRevLett.113.046401,PhysRevLett.116.186402}, and the ring-shaped nodal lines can be gapped \cite{PhysRevB.93.201114,PhysRevB.94.155121} or split into Weyl fermions \cite{PhysRevX.5.011029,PhysRevB.94.165201}. From this point of view, hunting for materials hosting versatile topological particles is of particular interests.

In this work, we propose that TaS is a new type of material
that hosts topological nodal lines, Weyl fermion states and unique fermion states with triply
degenerate crossing points near the Fermi level simultaneously.
There will be six pairs of WPs near the mirror plane $k_z=0$.
Every pairs of WPs possess opposite chirality (+1 or -1) and each WP
can be viewed as a singular point of Berry curvature or
``magnetic monopole" in momentum space \cite{PhysRevB.83.205101}.
Since TaS owns unique $D_{3h}$ symmetry, all WPs are located at the same energy.
Along the $\Gamma-A$ line in the BZ, four TDPs are presented in the vicinity of the Fermi energy. These TDPs are protected by the $C_{3}$ rotation
symmetry along $\Gamma-A$ direction. The coexistence of topological nodal lines, WPs and TDPs
states in TaS makes it a potential candidate to study the interplay between different types of topological fermions.

\section{MODELS AND METHODS}
\begin{figure}[ptbh]
\centering \includegraphics[width=1.\columnwidth]{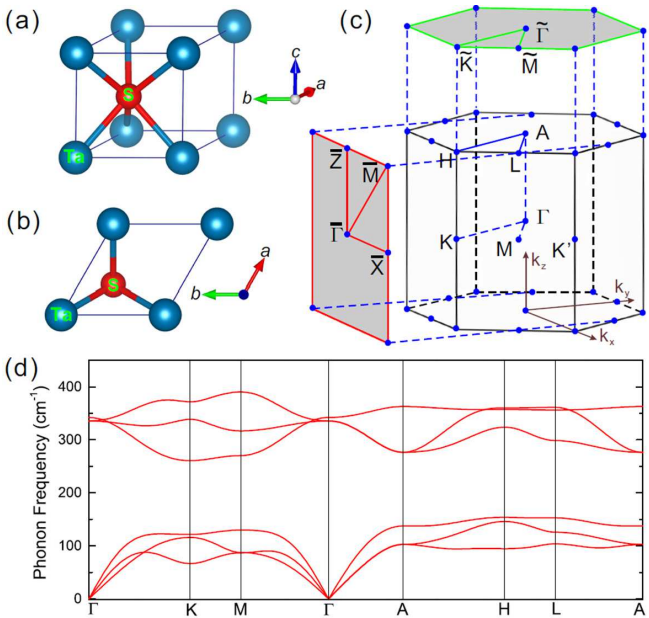} \caption{The overview (a) and top view (b) of the unitcell of the TaS single crystal. (c) The bulk Brillouin zone and its projection onto the (001) and (010) surfaces. (d) The phonon dispersion curves for TaS.}
\label{fig1}
\end{figure}

As shown in Fig. \ref{fig1}, TaS has a hexagonal
lattice structure \cite{1088}, with the same as Ta(Nb)N \cite{813943,91018,90777} and ZrTe \cite{ja004164r} which host same symmetry $D_{3h}^{1}$ with space group $P\bar{6}m2$ (No. 187). Ta and S atoms are located in $1a(0,0,0)$ and $1d(\frac{1}{3},\frac{2}{3},\frac{1}{2})$ Wyckoff positions, respectively.
The optimized lattice constants are $a=b=3.273$ \r{A} and $c=3.337$ \r{A}, which reasonably agree with experimental datas $a=b=3.280$ \r{A} and $c=3.145$ \r{A}.  To further prove the structural stability of
the optimized crystal, we calculate the phonon dispersions by using the frozen phonon method \cite{PhysRevLett.78.4063} as implemented in the PHONOPY code \cite{PhysRevB.78.134106}, as shown in Figure. \ref{fig1}(d). No imaginary
frequencies are observed throughout the whole BZ
in phonon dispersions, indicating its dynamically structural
stability. Therefore,
in this work, we adopt the relaxed lattice parameters for all the consequential calculations.

To explore the electronic structures of TaS, we perform the band structure calculation by using the Vienna $ab$ $initio$ simulation package (VASP) \cite{PhysRevB.54.11169} within the generalized gradient approximation (GGA) in Perdew-Burke-Ernzerhof (PBE) \cite{PhysRevLett.77.3865} type and the projector augmented-wave (PAW) pseudopotential \cite{PhysRevB.50.17953}. The kinetic energy cutoff is set to be 560 eV for wave-function expansion, and the $k$-point grid is sampled by sums over $12\times12\times12$ \cite{PhysRevB.13.5188}.
For the convergence of the electronic self-consistent calculations, the total energy difference criterion is set to $10^{-8}$ eV. The crystal structure is fully relaxed until the residual forces on atoms are less than 0.01 eV/\AA.
In order to investigate the projected surface states and Fermi surfaces, a  tight-binding model Hamiltonian based on the maximally localized Wannier functions (MLWF) method \cite{PhysRevB.56.12847,PhysRevB.65.035109} has been constructed. Then, we apply an iterative method \cite{016,009} implemented in the software package WannierTools \cite{tool} to obtain the surface Green's function of the semi-infinite system and the imaginary part of the surface
Green's function is the local density of states (LDOS) at the surface.

\section{RESULTS AND DISCUSSIONS}

\subsection{Overview of electronic structures of the bulk material}
\begin{figure*}[ptbh]
\centering \includegraphics[width=2.\columnwidth]{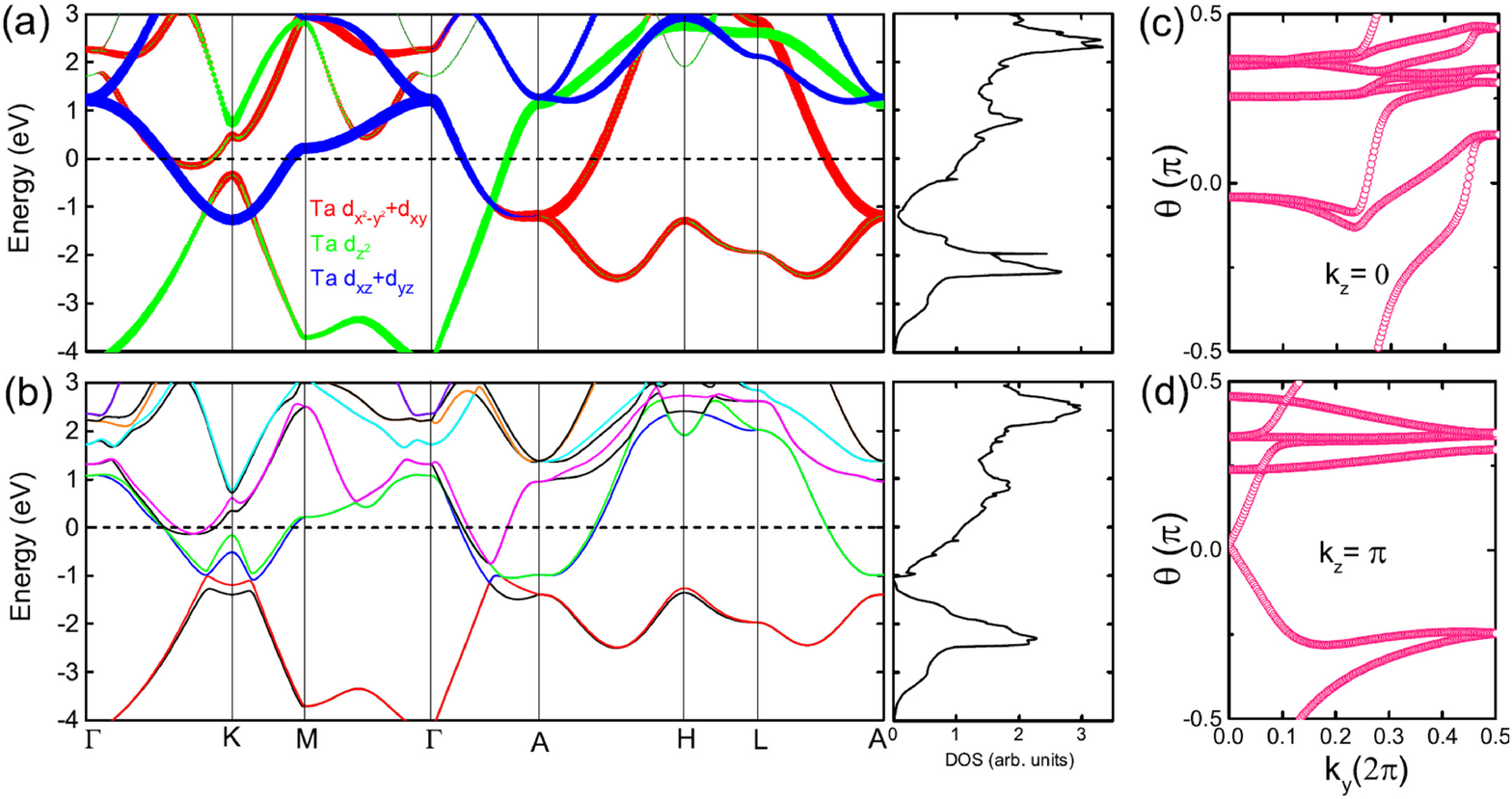} \caption{
Band structures and density of states (DOS) of TaS withouth (a) and with (b) the inclusion of SOC. The pink circles denote the evolution of Wannier centers for TaS along $k_y$ in the (c) $k_z = 0$ and (d) $k_z=\pi$ planes.}
\label{fig2}
\end{figure*}
The calculated band structures excluding the spin-orbit coupling (SOC) is shown in Fig. \ref{fig2}(a).
Near the Fermi level, the valence and conduction bands are mainly attributed
by the $d$ orbitals of Ta. In the absence of SOC, there is a band inversion around the $K$ point between $d_{x^2-y^2}+d_{xy}$ and $d_{xz}+d_{yz}$ of Ta atom, which
leads to a gapless nodal ring centering at $K$ point in the horizontal $k_z=0$ mirror plane protected by $\sigma_h$ symmetry. The similar nodal ring band topology is observed in ZrTe \cite{PhysRevX.6.031003,PhysRevB.94.165201} and MoC \cite{027102}, but is
missing in TaN \cite{PhysRevX.6.031003}. Since the three materials share the same space group with TaS, the existence of nodal ring is not simply determined by the crystal symmetry.
Unlike the character of inverted band structures around the $K$ point, we find a band crossing between one nondegenerate band composed by $d_{z^2}$ orbital of Ta atom and a double degenerate band composed by the combination of $d_{x^2-y^2}$ and $d_{xy}$ orbitals of Ta atom along the $\Gamma-A$ line in the reciprocal lattice. The crossing point is a TDP protected by the $C_{3v}$ symmetry in the absence of SOC.
Since TaS hosts the heavy element Ta, the SOC can not be simply ignored.
By introducing the SOC, due to the lack of inversion
symmetry, the energy bands happen to split in momentum space, as demonstrated in Fig. \ref{fig2}(b). From Fig. \ref{fig2}(b), one can find out that, the nodal ring centering at $K$ point is gapped and the original TDP splits into four discrete crossing points along the $\Gamma-A$ symmetric line.
Considering that the electronic structures
within the $k_z = 0$ and $k_z = \pi$ planes can be decoupled and regarded as individual 2D subsystems with time-reversal symmetry, we can give a well defined $\mathds{Z}_2$ number to determine the band topology
for all the occupied bands. Since TaS has no inversion symmetry,
we use the method of the evolution of Wannier charge
centers (WCCs) \cite{PhysRevB.83.235401,PhysRevB.83.035108,PhysRevB.89.115102} to calculate the $\mathds{Z}_2$ number by counting how many times the evolution lines of the Wannier centers cross an arbitrary reference line. In Fig. \ref{fig2}(c) and \ref{fig2}(d), we find that both $k_z=0$ and $k_z=\pi$ planes are topologically nontrivial with $\mathds{Z}_2 = 1$.

\subsection{Nodal lines and triply degenerate points}
In order to understand the role of SOC in the nodal ring gapping and TDP splitting, we calculated the detailed band structures of TaS in the absence and presence of SOC within specific BZ regions.

\begin{figure}[ptbh]
\centering \includegraphics[width=1\columnwidth]{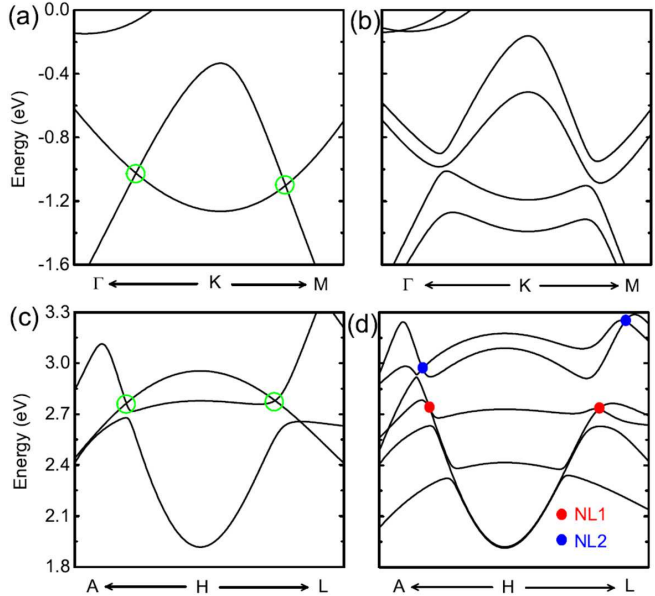} \caption{Band structures around $K$ point in the absence of (a) and presence of (b) SOC.
Band structures around $H$ point in the absence of (c) and presence of (d) SOC. In the panels, the  locations of the ring-shaped nodal lines are indicated by the green circles, red (NL1) and blue (NL2) dots.}
\label{fig3}
\end{figure}
\begin{figure}[ptbh]
\centering \includegraphics[width=1\columnwidth]{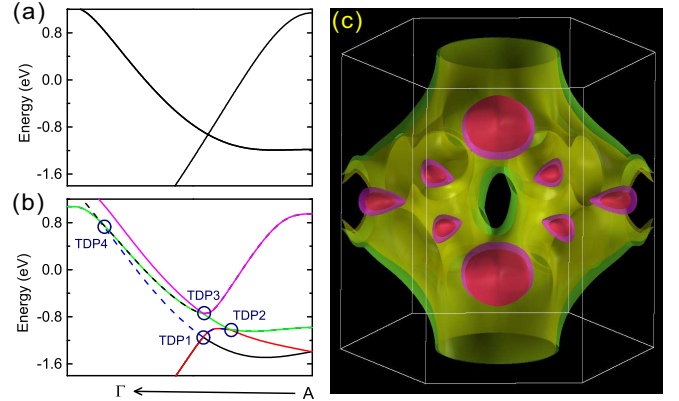} \caption{
Band structures along $\Gamma-A$ without (a) and with (b) the SOC.
(c) Overwiew view of the Fermi surface of TaS with $E = 0$ eV.}
\label{fig4}
\end{figure}

The enlarged band structures around the $K$ and $H$ points are shown in Fig. \ref{fig3}. In the absence of SOC, there are two nodal lines appearing on the
 mirror plane of $k_z=0$ and $k_z=\pi$, respectively. The nodal line on the $k_z=0$ plane
is located about 1.1 eV below the Fermi level centering at the $K$ point, while the
other one on the $k_z=\pi$ plane is located about 2.8 eV above the Fermi level centering at the $H$ point, as shown in Fig. \ref{fig3}(a) and \ref{fig3}(c).
By turning on the SOC, the nodal ring around the $K$ point
is fully gapped, as shown in Fig. \ref{fig3}(b). However, different from the $K$ point, the nodal ring around the $H$
point still preserves and splits into two nodal rings marked by red and blue dots respectively in Fig. \ref{fig3}(d) in the presence of SOC. The preserving of nodal ring under SOC is strictly protected by the mirror symmetry $\sigma_h$.
Along the $\Gamma-A$ direction, there are four triply degenerate crossing points
emerging near the Fermi level due to the band splitting induced by SOC, as shown in Fig. \ref{fig4}. Due to the the time-reversal symmetry, these TDPs appear in pairs and are protected by threefold rotation symmetry $C_3$ and vertical mirror symmetry $\sigma_v$.
The four TDPs located within a relatively large energy scale ranging from -1.141 eV to 0.812 eV are a direct result stemming from the large SOC in the heavy element Ta.
For a more direct view of the degeneracy and separation of energy bands, the 3D Fermi surface at $E=0 $ eV is presented in Fig. \ref{fig4}(c).

\subsection{Weyl nodes}
\begin{figure}[pht]
\centering \includegraphics[width=1\columnwidth]{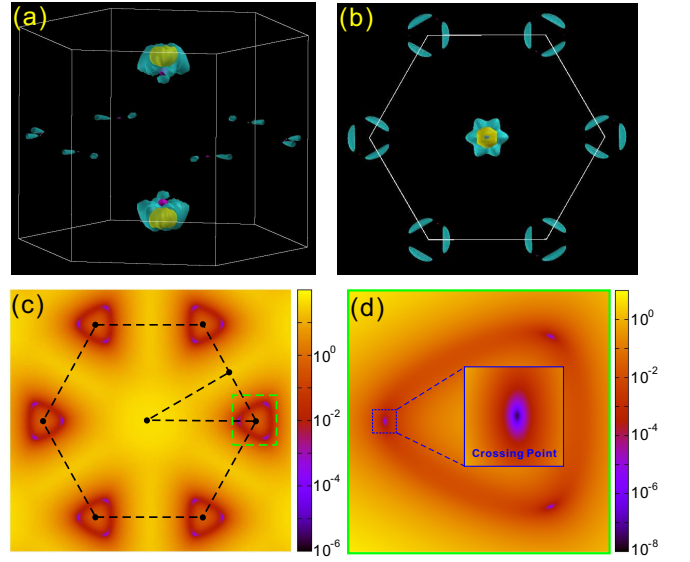} \caption{
 (a) The overview and (b) the top view of the isoenergy surface at $E=-1.026$ eV. (c) The energy difference between the blue and red bands in
Fig. \ref{fig2} in the $k_z=0.01006\times\frac{2\pi}{c}$ plane, and the energy difference in the green box is enlarged and demonstrated in panel (d), the contour box in unit of eV indicates the energy difference at the crossing points are zero.}
\label{fig6}
\end{figure}

Although the nodal ring preserves around the $H$ point on the $k_z=\pi$ plane, the nodal ring centering at the $K$ point on the $k_z=0$ plane is fully gapped, and the gapped nodal ring may produce a transition into hidden Weyl points. Generally, it is difficult to identify WPs within the momentum space because the WPs are not protected by any additional crystalline symmetry
besides the lattice translation and are possibly located at general $\mathbf{k}$ points. However, detailed electronic structure calculations give us some clues.
As shown in Fig. \ref{fig6}(a) and \ref{fig6}(b), by tuning the isoenergy reference into the opened gap, one can observe vanishing connections between the three-fold rotational symmetric arcs centering at the $K$ point, which indicates the DOS of the connections become zero. However, by further calculations of the energy gap between blue and red bands in Fig. \ref{fig2}(b),
we find that the arc connections are simultaneously bands crossing points, which are demonstrated in Fig. \ref{fig6}(c) and \ref{fig6}(d).
As a consequence,
we find six pairs of zero mass band crossing points off the $k_z=0$ plane within the gap opened in the original nodal ring, and one pair of
typical crossing points are located at $(0.27449,0.27449,\pm0.01006)$ in unit of reciprocal lattice vectors marked as W+ and W-, respectively.

Following, we will check that whether these band crossing points are WPs or not. As an important clue, WPs can be viewed as monopoles with opposite chirality corresponding to the sources and sinks of Berry curvature.
Besides, the integral of the Berry curvature on any closed surface in the BZ equals the total chirality of the WPs enclosed by the given surface, which will
help us chase down and check WPs quickly. Therefore,
to further verify the connecting points host monopole feature and chirality, we apply Kubo formula to
calculate the Berry curvature by the tight-binding
Hamiltonian based on the MLWF \cite{RevModPhys.82.1959,PhysRevB.74.195118},
\begin{equation}
\hat{v}_{\alpha(\beta\gamma)}=\left(  i/\hbar\right)  \left[  \hat{H},\hat
{r}_{\alpha(\beta\gamma)}\right]  =\frac{1}{\hbar}\frac{\partial\hat
{H}(\mathbf{k})}{\partial k_{\alpha(\beta\gamma)}},%
\end{equation}
\begin{equation}
\Omega_{n,\alpha\beta}(\mathbf{k})=-2\operatorname{Im}\sum\limits_{m\neq
n}\frac{\langle u_{n\mathbf{k}}|\hat{v}_{\alpha}|u_{m\mathbf{k}}\rangle\langle
u_{m\mathbf{k}}|\hat{v}_{\beta}|u_{n\mathbf{k}}\rangle}{\left[  \omega
_{m}(\mathbf{k})-\omega_{n}(\mathbf{k})\right]  ^{2}},%
\end{equation}
where $\hat{v}_{\alpha(\beta\gamma)}$ is the velocity operator with $\alpha,\beta,\gamma=x,y,z$, $\omega_{m}(\mathbf{k})=\varepsilon_{n}(\mathbf{k})/\hbar$,
$\varepsilon_{n}(\mathbf{k})$ and $u_{n\mathbf{k}}$ are the eigenvalue and
eigenvector of the Hamiltonian $\hat{H}(\mathbf{k})$, respectively.
Finally, we obtain the Berry curvature in momentum space for a
given band $n$ by
\begin{equation}
\Omega_{n,\gamma}(\mathbf{k})=\epsilon_{\alpha\beta\gamma}\Omega
_{n,\alpha\beta}(\mathbf{k}),
\end{equation}
where $\epsilon_{\alpha\beta\gamma}$ is the Levi-Civita tensor.
By summing over all occupied valence bands, we give the Berry curvature
vector $
\mathbf{\Omega=}(\Omega_{x},\Omega_{y},\Omega_{z})$.
\begin{figure}[pb]
\centering \includegraphics[width=1\columnwidth]{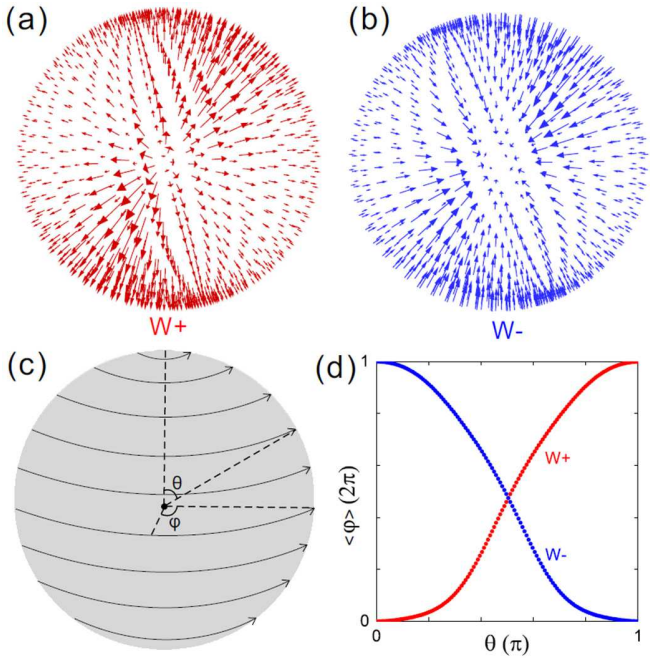} \caption{The
Berry curvature around the (a) W+
and (b) W-, respectively. (c) schematic illustration of the integration paths used to calculate topological charges of Weyl points. (d) Evolution of the Wannier charge centers around the sphere enclosing W+ (red) and W- (blue).}
\label{fig7}
\end{figure}
Next, we will examine the monopole feature of the crossing points.
Since a pair of WPs with opposite chirality
represent the source and sink of the Berry curvature, we calculate the Berry curvature distribution in momentum space. From Fig. \ref{fig7}(a) and \ref{fig7}(b), we can clearly see the Berry curvature around W+ and W- is divergent and convergent, respectively, which indicates the W+ and W- act as the feature of the source and sink of the Berry curvature.

In order to confirm the chirality of the crossing points, we need get the
flux of Berry curvature through a surface enclosing them. The corresponding flux of Berry curvature is computed by discretizing a closed sphere parametrized by angles $\theta$ and $\varphi$ into 1D-loops, as shown in
Fig. \ref{fig7}(c). Then, We calculate the WCCs on longitudinal
loops around a sphere enclosing the crossing point. The sum of the WCCs on a loop $\mathcal{C}$ with longitudinal $\theta$ angle corresponds to the accumulated Berry phase along the loop, or equal to the average position $\left\langle \varphi\right\rangle$ of the charge on the loop.
It can be
determined from the Wilson loop $\mathcal{W}(\mathcal{C})$ as
\begin{equation}
\varphi(\theta)\mathcal{=}\operatorname{Im}\{\ln[\det\mathcal{W}%
(\mathcal{C)}]\},
\end{equation}%
where
$
\mathcal{W}(\mathcal{C})=\prod\limits_{i=0}^{L-1}M^{(\mathbf{k}_{i,}%
\mathbf{k}_{i+1})}
$, $\left[  M^{(\mathbf{k}_{i,}\mathbf{k}_{i+1})}\right]_{mn}   =\langle
u_{m\mathbf{k}_{i}}|u_{n\mathbf{k}_{i+1}}\rangle$ are the overlap matrices
along the loop $\mathcal{C}$.
Since 1D loops cover a closed surface, the center of charge $\left\langle \varphi\right\rangle$ can only shift by an integer multiple of $2\pi$ when $\theta$ varies from 0 to $\pi$.
This multiple is equal to the chirality $\mathcal{C}$ of the crossing point enclosed in the sphere.
Therefore, we apply the method to a sphere enclosing W+ or W-. As shown in Fig. \ref{fig7}(d), when the angle $\theta$ varies from 0 to $\pi$, the average position of the Wannier centers $\left\langle \varphi\right\rangle$ is shifted by $2\pi$ or $-2\pi$, which indicates W+ and W- are indeed a pair of
 WPs with chirality $\mathcal{C}=+1$ and $\mathcal{C}=-1$, respectively.

\begin{figure}[ptbh]
\centering \includegraphics[width=1\columnwidth]{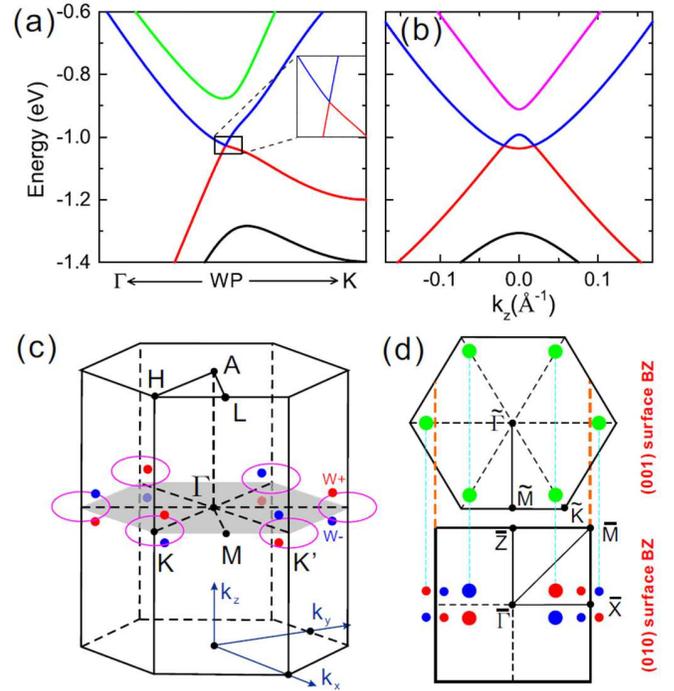} \caption{
The detail and overview of the band dispersions around the WPs are demonstrated in panel (a) and panel (b). (c)
The original gapless nodal rings in the absence of SOC are denoted by the pink rings, and the emergent WPs in the presence of SOC in the gapped nodal rings are illustrated by the dots and their chirality are distinguished by the red (+1) and blue (-1) dots.
(d) The projected WPs onto (001) and (010) surfaces, respectively. The green large dots indicate projections of a pair of WPs with opposite chirality, and the chirality becomes indistinguishable. The red and blue large dots indicate two projected WPs with
same chirality, while the small dots indicate the single projected
WP.
}\label{fig5}
\end{figure}

Around the WPs (W+ and W-), we plot the energy dispersions, as
shown in Fig. \ref{fig5}(a) and \ref{fig5}(b). Since the TaS hosts high symmetry
$D_{3h}$ including $C_3$ rotation and mirror symmetry et al, which results in
all WPs are located at the same energy (1.026 eV below the Fermi level).
By utilizing above comprehensive techniques, we obtain at once the positions and chirality of all six pairs of WPs in the first whole BZ, which are clearly presented in Fig. \ref{fig5}(c).
All these WPs are located near the $K$ point and its time reversa $K'$ point
and are off the $k_z=0$ mirror plane.
The Fig. \ref{fig5}(d) displays the projection of all WPs onto the top and slide surfaces. When these WPs are projected onto top (001) surface, each pair of WPs with
opposite chirality are projected onto a same position labeled by large green dots.
Whereas on the protected (010) surface, the WPs with different chirality are
projected onto a different position. The large and small dots represent
the projections of two WPs with same chirality and a single WP, respectively.

\subsection{Surface states and Fermi arcs}

\begin{figure}[ptbh]
\centering \includegraphics[width=1\columnwidth]{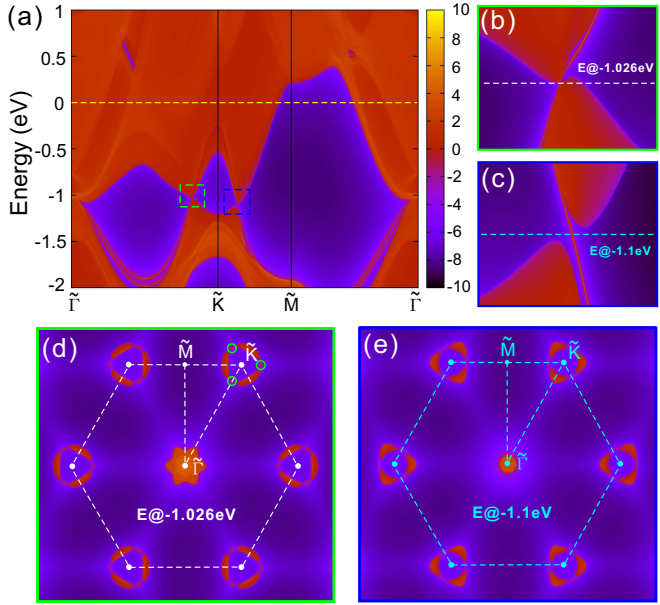} \caption{(a) The projected surface density of states for the (001) surface.
(b) and (c) The enlarged surface states in (a).
(d) and (e) The Fermi surface of the projected (001) surface in the whole BZ with $E=-1.026$ eV and $E=-1.1$ eV, respectively.}
\label{fig8}
\end{figure}
Since different projections bring different expressions of chirality of WPs, we calculate the projected surface states and Fermi surfaces (Fermi arcs) onto the (001) and (010) surfaces.
The (001)-surface states and Fermi surfaces are shown in Fig. \ref{fig8}.
Around the $\tilde{K}$ point, there is a analogous nodal ring but with a bulk gap
about 30 meV due to the spin-orbit interaction along the $\tilde{K}-\tilde{M}$ direction, as shown in Fig. \ref{fig8}(c).
In Fig. \ref{fig8}(b), a Weyl node is exactly located at
$-1.026$ eV on the $\tilde{\Gamma}-\tilde{K}$ line, which mainly roots in the contribution of the projected WP. Around the symmetric $\tilde{K}$ or $\tilde{K}^\prime$ point in the BZ, there are three pairs of projected WPs existing, which will generate threadlike Fermi arcs, as clearly presently in Fig. \ref{fig8}(d). Each green circle
represents the projections of two WPs with opposite chirality, which interconnect
by the Fermi arcs.
Moreover, since the TDPs are located on the $\Gamma-A$ line,
they are projected on the $\tilde{\Gamma}$ point surrounded
by the bulk state projections.

\begin{figure}[pt]
\centering \includegraphics[width=1\columnwidth]{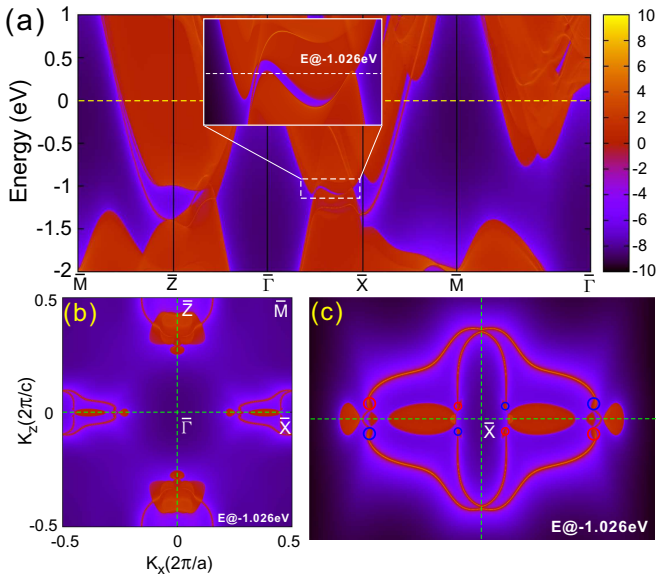} \caption{(a) The projected surface density of states for the (010) surface. (b) The Fermi surface  of the projected (010) surface with $E=-1.026$ eV. The Fermi surface around the $\bar{X}$ point in (b) is
enlarged in (c), and where the projected WPs carrying opposite chirality are indicated by the red and blue circles, respectively.}
\label{fig9}
\end{figure}

For the (010) surface states and Fermi surfaces, which are presented in Fig. \ref{fig9}. From the Fig. \ref{fig9}(a), one can clearly see that
there is a surface Dirac cone at $\bar{Z}$ point, which is a result of $Z_2=1$ of the $k_z=\pi$ plane. The two branches of the surface
Dirac cone connect the bulk states of the conduction and valence bands, respectively. In order to observe the fancy Fermi arcs, we plot the Fermi surface
at $E=-1.026$ eV, as shown in Fig. \ref{fig9}(b) and \ref{fig9}(c).
In the Fig. \ref{fig9}(c), the projected WPs and Fermi arcs
are clearly presented, and each Fermi arc connects the projections of the WPs with opposite chirality. The red and blue large circles represent the
projections of two WPs with the same chirality and have a monopole charge
$+2$ and $-2$, respectively. While the red and blue small circles
represent a single projected WP and have a monopole charge
$+1$ and $-1$, respectively. This is consistent with
the description of Fig. \ref{fig5}(d).
In addition,  the TDPs located at the $\Gamma-A$ line in the bulk band are projected onto the $\bar{\Gamma}-\bar{Z}$ path, and they can not be detached from other surface states and are buried in the projected bulk states, which is mainly
due to the breaking of the $C_3$ and vertical mirror plane $\sigma_v$ symmetry on (010) slide surface.

\subsection{Lifshitz transition}
\begin{figure}[pb]
\centering \includegraphics[width=1\columnwidth]{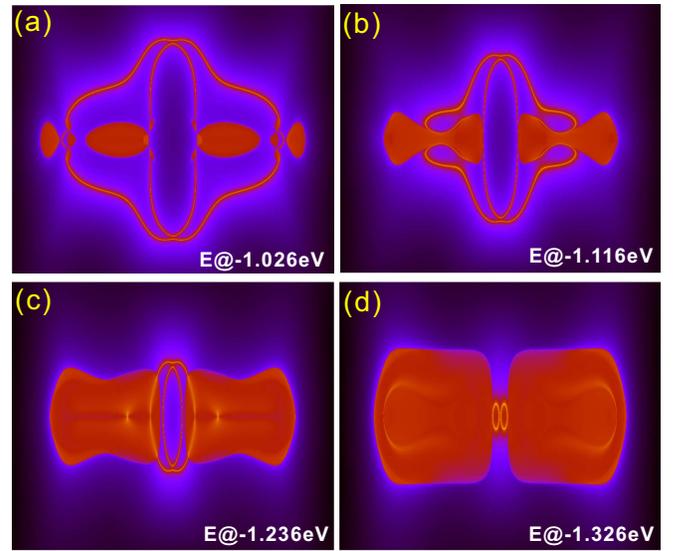} \caption{(b) The evolution of the isoenergy surface contours around the $\bar{X}$ point of the projected (010) surface with respect to the chemical potentials are demonstrated at (a) $E=-1.026$ eV, (b)
$E=-1.116$ eV, (c) $E=-1.236$ eV, (d) $E=-1.326$ eV.}
\label{fig10}
\end{figure}
Since the connections between WPs carrying opposite chirality preserve on the (010) projection surface, and each pair of different chiral WPs are located very close to each other in momentum space, they can inevitably hybridize together. By adjusting chemical potential appropriately, one can expect a topological change in the band contours, which is known as the Lifshitz transition \cite{PhysRevLett.115.166602,PhysRevB.92.075115,4974185,00849}.
In Fig. \ref{fig10}, we present the Lifshitz transition by changing the contour
energy. At $E=-1.026$ eV, Fermi arcs connecting the projected WPs with opposite
chirality are clearly shown in Fig. \ref{fig10}(a), which has been
introduced in detail on the above. As the contour energy decreases,
the Fermi arcs disappear and the topological properties will change.
The projected surface bands will close to each other, then form two
concentric ellipses, finally develop into two ellipses side by side.

\section{SUMMARY AND CONCLUSIONS}

In conclusion, in this paper we report that, TaS is a stable topological semimetal hosting versatile quasiparticle excitations such as topological nodal rings, Weyl points and three-fold degenerate points protected by the rotation
symmetry $C_3$ and vertical mirror symmetry $\sigma_v$ simultaneously in the presence of SOC.
Based on the DFT calculations, we study
the topologically nontrivial properties in TaS both in bulk states and projected surface states. We find the nodal ring centering at $H$ point of the moment space remain closed to meet the requirement of the horizontal mirror symmetry $\sigma_h$ and the nodal ring centering at $K$ point is fully gapped by introducing the SOC.
In the gapped nodal ring, there are six pairs of WPs associating with the point group $D_{3h}$
and locating at the same energy
in the first BZ. Chirality and monopole feature of the WPs are verified by the calculation of the Berry curvature, and the characteristic Fermi arcs are found on the particular projected
surfaces, which connects the projections of WPs with opposite chirality.
By the projection onto the (010) surface and adjustment of chemical potential, a Lifshitz transition is also observed.
The coexistence of versatile topological excitations brings extraordinary properties such as drumhead surface states and negative magnetoresistance effects. In light of these properties of the electronic band structures and surface states in TaS,
we provide a guiding principle to search the particular
topological semimetal, and establish an experimental achievable material platform for studying the unique topological phenomena in solids.

\section*{ACKNOWLEDGMENTS}
This work was supported by NSFC Grants No. 11504366 and NBRPC Grants
No. 2015CB921503 and 2016YFE0110000.

\bibliographystyle{apsrev4-1}
\bibliography{refer}

\end{document}